\documentclass[aps,pre,preprint,showkeys,superscriptaddress]{revtex4} 

\usepackage[T1]{fontenc}
\usepackage{amsmath,amssymb}
\usepackage{graphicx}
\usepackage{bm}
\usepackage{enumitem}
\usepackage{xcolor}
\definecolor{blue}{rgb}{0.0,0.0,0.0}

\begin{document}

\title{Modelling intermittent anomalous diffusion with switching fractional
Brownian motion}

\author{Micha{\l} Balcerek} 
\author{Agnieszka Wy{\l}oma{\'n}ska} 
\author{Krzysztof Burnecki}
\affiliation{Faculty of Pure and Applied Mathematics, Hugo Steinhaus Centre, Wroc{\l}aw University of Science and Technology, 50-370 Wroc{\l}aw, Poland}
\author{Ralf Metzler} 
\affiliation{Institute of Physics \& Astronomy, University of Potsdam, 14476
Potsdam, Germany}
\affiliation{Asia Pacific Centre for Theoretical Physics, Pohang 37673, Republic
of Korea}
\email{rmetzler@uni-potsdam.de}
\author{Diego Krapf}
\affiliation{Department of Electrical and Computer Engineering, Colorado State
University, Fort Collins, Colorado 80523, USA}
\email{diego.krapf@colostate.edu}

\date{\today}

\begin{abstract}
The stochastic trajectories of molecules in living cells, as well as the
dynamics in many other complex systems, often exhibit memory in their path
over long periods of time.  In addition, these systems can show dynamic
heterogeneities due to which the motion changes along the trajectories. Such
effects manifest themselves as spatiotemporal correlations. Despite the broad
occurrence of heterogeneous complex systems in nature, their analysis is
still quite poorly understood and tools to model them are largely missing. We
contribute to tackling this problem by employing an integral representation
of Mandelbrot's fractional Brownian motion that is compliant with varying
motion parameters while maintaining long memory. Two types of switching
fractional Brownian motion are analysed, with transitions arising from a
Markovian stochastic process and scale-free intermittent processes. We obtain
simple formulas for classical statistics of the processes, namely the mean
squared displacement and the power spectral density. Further, a method to
identify switching fractional Brownian motion based on the distribution of
displacements is described. A validation of the model is given for experimental
measurements of the motion of quantum dots in the cytoplasm of live mammalian
cells that were obtained by single-particle tracking.
\end{abstract}

\keywords{heterogeneous diffusion; stochastic processes; single-molecule
tracking; ergodicity breaking, long memory, cytoplasm}

\maketitle

\section{Introduction \label{sec:intro}}

The statistical analysis of particle trajectories recorded with
single-particle tracking has revolutionised the field of cellular biophysics
\cite{levi2007exploring,manzo2015review,barkai2012strange,hofling2013anomalous,
krapf2019strange}. To name a few representative
examples, exquisite information is found on lipid membranes
\cite{dietrich2002relationship,knight2009single,campagnola2015superdiffusive},
receptors \cite{manzo2015weak,metz2019temporal,mosqueira2020antibody},
ion channels \cite{weigel2011ergodic,akin2016single,he2016dynamic},
nucleic acids \cite{bronstein2009transient,moon2019multicolour},
filaments \cite{ruhnow2011tracking}, and organelles
\cite{nixon2016increased,speckner2018anomalous,korabel2021local}. Further,
synthetic particles can be used as probes to study cellular rheology
\cite{weihs2006bio,etoc2018non,sabri2020elucidating}. Beyond intracellular
dynamics, individual stochastic trajectories are studied in a large variety of
fields, including the motion of flagellated organisms \cite{berg2000motile},
larvae \cite{sims2019optimal}, marine predators \cite{hays2012high},
and birds \cite{vilk2022unravelling,vilk2022ergodicity},
as well as the fluctuations in financial markets
\cite{bouchaud2005subtle,scalas2006application} and percolation in porous
materials \cite{edery2010particle,weigel2012obstructed,wu2020nanoparticle}. All
these complex systems can be characterised in terms of similar
statistics, such as the second moment, the distribution of
displacements, temporal correlations, and spectral components
\cite{metzler2014anomalous,krapf2015mechanisms,krapf2018power}. Frequently,
trajectories in complex systems exhibit anomalous diffusion defined by a
non-linear mean squared displacement (MSD). In particular, the MSD of a
process $X(t)$ is often observed to scale as a power-law in time, i.e.,
$\langle X^2(t) \rangle\propto t^\alpha$, where the angular brackets denote
an ensemble average. The parameter $\alpha$ is the anomalous diffusion
exponent and it classifies the process as being subdiffusive when $\alpha<1$
and superdiffusive when $\alpha>0$. In contrast, Brownian motion has a linear
MSD, $\alpha=1$, and ballistic, wave-like motion corresponds to $\alpha=2$.

Several models have been successfully employed to describe particle motion
within the framework of anomalous diffusion \cite{metzler2014anomalous}. From
single-particle trajectories, anomalous diffusion processes
can be distinguished by complementary statistical observables
\cite{metzler2014anomalous} enabling the construction of decision
trees \cite{yazmin}, or by Bayesian as well as deep learning
approaches \cite{burrage,thapa,munoz2021objective,henrik,henrik1,janusz,
gorka}. Among the anomalous diffusion processes, the
continuous time random walk (CTRW)  with  scale-free sojourn times
\cite{montroll1965random,scher1973stochastic,scher1991time} and
fractional Brownian motion (FBM) with long-ranged temporal correlations
\cite{kolmogorov1940wienersche,mandelbrot1968fractional} are the most
widespread. In the CTRW model, a particle performs a random walk in
which the waiting times between jumps are stochastic with a probability
density function (PDF) $\psi(t)$. When the PDF of the waiting times
has the scale-free form $\psi(t)\sim t^{-1-\beta}$ with $0<\beta<1$,
the mean waiting time diverges and the motion follows a subdiffusive
pattern. The scale-free CTRW has many counterintuitive properties because
the process is non-stationary \cite{metzler2014anomalous}. FBM describes a
self-similar process with stationary, power-law correlated, and Gaussian
increments, of which Brownian motion constitutes a special case. FBM is
particularly useful in modelling anomalous transport with memory effects
\cite{szymanski2009elucidating,magdziarz2009fractional,sadegh2017plasma}.

While many correlated motions are well described by FBM, in multiple instances
it is found that the increments are not Gaussian \cite{lampo2017cytoplasmic,
he2016dynamic,jeon2016protein,sabri2020elucidating,Balcerek2023}. Further, in other
striking observations, correlated motions exhibit non-ergodicity,
that is, the nonequivalence between the ensemble-averaged MSD
and the time-averaged MSD for sufficiently long trajectories
\cite{weigel2011ergodic,jeon2011vivo,tabei2013intracellular}. Importantly,
Gaussianity and ergodicity are hallmarks of unconfined FBM
\cite{deng2009ergodic}. The underlying key reasons for these complex
effects, non-Gaussianity in particular, in FBM-like correlated
processes are heterogeneities that arise both from trajectory to
trajectory and, even, within individual trajectories. Notably, it
is often observed that the state of a system can change in time
due to dynamic interactions or a shift in the properties of the
environment. Heterogeneous dynamics have been identified in trajectories
from proteins and lipids in the plasma membrane \cite{choquet2013dynamic,
he2016dynamic,weigel2013quantifying,jeon2016protein,sikora2017elucidating,
weron2017ergodicity}, vesicles that move along cytoskeleton
filaments \cite{arcizet2008temporal}, intracellular transport of
endosomes and lysosomes \cite{fedotov}, and DNA-binding proteins
\cite{loverdo2009quantifying}. \textcolor{blue}{In Fig.~\ref{Trajectories}a
we show three trajectories of quantum dots recorded within live HeLa cells
\cite{sabri2020elucidating}, as a visual example for experimental trajectories,
in which the state changes within individual trajectories.
On top of these examples, other fields, in which} regime
changes play a significant role within individual trajectories with
anomalous dynamics, include biomedical signals \cite{andreao2006ecg},
speech \cite{khanagha2014phonetic}, traffic flows \cite{cetin2006short},
econometrics \cite{janczura2013goodness,lux2010forecasting}, ecology
\cite{edelhoff2016path}, solar activity \cite{stanislavsky2009farima},
and river flows \cite{vasas2007two}.

Despite the large number of experimental systems unveiling anomalous transport
that exhibits transitions between diffusive states, their computational
and theoretical analyses are mostly missing. This type of analysis is
critical to understanding spatiotemporal kinetics in heterogeneous complex
systems. One of the main issues is the lack of tools to simulate processes
that continuously maintain long-range correlations after a regime change
is encountered. The standard procedure relies on the assumption that the
process encounters a renewal at each regime change, i.e., the memory is lost
when the state changes. Alternatively, subordination schemes can be used
for the study of immobilisations. However, what is missing is a tool that
allows for computational studies of switching long-range correlated motion.

In this article, we employ a modified stochastic integral representation
to simulate FBM trajectories with discretely switching parameters. Our
representation is based on L{\'e}vy's formulation \cite{levy1953random} and
it is generalised to having time-dependent diffusion coefficient $D$ and
anomalous diffusion exponent $\alpha$. In particular, $D$ and $\alpha$ are
considered to be stochastic processes, so that the trajectory switches between
different states as function of time. We study two specific processes; in the
first case, the dwell times in each state are exponentially distributed and,
in the second, a state has dwell times with a heavy-tailed distribution. The latter
yields a process that is aging and non-ergodic. The numerical simulations
are analysed in terms of the MSD and the power spectral density (PSD). 
Closed-form asymptotic formulas are obtained for both analyses. Our results
are compared to those obtained from the experimental trajectories of quantum
dots in the cytoplasm of mammalian cells \cite{sabri2020elucidating}, which
is a well-characterised system showing correlated increments with random
switching between two states.

\section{Methods}

\subsection{Numerical simulations}

The classical FBM $B_H(t)$ 
is a continuous process with autocovariance function
\cite{mandelbrot1968fractional}
\begin{equation}
\label{autocorrFBM}
\left<B_{H}(t)B_{H}(s)\right> =D\left(t^{2H}+s^{2H}-|t-s|^{2H}\right),
\end{equation}
where $H\in(0,1]$ is the Hurst exponent and the
generalised diffusion coefficient $D$ is a constant with units
$\mathrm{length}^2/\mathrm{time}^{2H}$. 
For $H=1/2$, the process becomes the standard Brownian motion $B(t)$, so $B_{1/2}(t)=B(t)$.
Eq.~\eqref{autocorrFBM} yields an
MSD of the form $\left<B_H^2(t)\right> =2Dt^{2H}$, which implies that the
anomalous diffusion exponent is $\alpha=2H$. 
The FBM is well-defined for all $t\in \mathbb{R}$. For $t\geq 0$, which is of our interest, the process can be approximated via 
L\'evy's formulation \cite{levy1953random,mandelbrot1968fractional} of non-equilibrated FBM in
terms of a Riemann-Liouville fractional integral, $B_H(t)= \sqrt{2DH} \int
^t_0 \; (t-s)^{H-1/2} dB(s)$.
Following our recently introduced process for time-dependent Hurst exponent
\cite{wang2023memory} we consider $D$ and $H$ to be explicitly time-dependent,
\begin{equation}
\label{defFBMSE}
X(t)= \int ^t_0 \sqrt{2D(s)H(s)} \; (t-s)^{H(s)-1/2} dB(s).
\end{equation}
To simulate switching FBM trajectories we use an Euler approximation
to discretise the integral.  Namely, we generate time series of Brownian
motion $B(t)$ increments and those of stochastically varying Hurst exponents $H(t)$ and
diffusivities $D(t)$, in an interval $[0,T]$. We then employ the discretised
integral \eqref{defFBMSE} to generate a switching FBM. The specifics of the
time series $H(t)$ and $D(t)$ depend on the process under investigation. In
the Results section, we present processes with two states where $H=0.1$,
$0.3$, $0.6$, or $0.8$, and $D=1$, $10$, or $100$. The dwell times in each
state are drawn from exponential (see Eq.~\eqref{dwellMarkov}) or Pareto 
(see Eq.~\eqref{dwellPowerLaw}) distributions. 
For exponential distributions, we employ mean dwell times $\tau=15$,
$25$, or $45$, and, for Pareto distribution, we use a scale parameter $t_0=15$
and shape parameter $\beta=0.7$. For each case, we generate 1,000 realisations
of 8,192 data points.

\subsection{MSD and PSD}

We characterise the diffusion processes in terms of two broadly used analyses,
the MSD and the PSD. Most typically, the MSD is evaluated as a time average
because it substantially augments the statistics. The time-averaged MSD is
defined as
\begin{equation}
\overline{\delta^2(\Delta,T)} = \frac{1}{T-\Delta}\int_0^{T-\Delta}
\left[X(t+\Delta)-X(t)\right]^2 dt,
\label{TAMSD}
\end{equation}
where $\Delta$ is the lag time and $T$ the measurement time. Further, an
ensemble average is performed over the time-averaged MSD, i.e., $\langle
\overline{\delta^2(\Delta,T)} \rangle$.

The PSD of a single-trajectory is defined as 
\begin{equation}
\label{1spec}
S(\omega,T)=\frac{1}{T} \left|\int^T_0 \exp(i\omega t)X(t)\; dt\right|^2, 
\end{equation}
where $\omega$ is the frequency. While for stationary processes, the PSD
is usually defined in the limit that $T$ approaches infinity, we employ a
more general definition where the spectral content explicitly depends on
both frequency and observation time, $S(\omega,T)$ \cite{krapf2018power}. As
with the MSD, the ensemble average of the single-trajectory PSD is computed,
i.e., $\langle S(\omega,T) \rangle$.

To simplify the notation, in the following we will refer to the
ensemble-averaged time-averaged MSD and the ensemble-averaged single trajectory
PSD, as the MSD and PSD, respectively.

\subsection{Quantum dot imaging and single-particle tracking}

Full experimental details were previously described
\cite{sabri2020elucidating}. Carboxylate functionalised quantum dots (Qdot
655 ITK, ThermoFisher, Waltham, MA) were incorporated into HeLa (human
cervical cancer) cells by bead loading. Cells were plated 36-48 h prior
to bead loading on 35 mm dishes (Delta T culture dish, Bioptechs, Butler,
PA), coated with 0.5\% matrigel (Corning Life Sciences, NY). Images were
acquired with an EMCCD camera at 10 frames/s on a custom-built microscope
equipped with an Olympus PlanApo 100x NA1.45 objective, and a CRISP ASI
autofocus system. During imaging, cells were maintained at 37 $^\circ$C and
the quantum dots were excited at 561 nm. Trajectories were extracted from
image stacks using the TrackMate ImageJ plugin.

\section{Results}

\subsection{Markovian switching between states}

We first consider a switching FBM with two states whose dwell times
are exponentially distributed. Thus, for each state,
\begin{equation}
\label{dwellMarkov}
\psi_i(t)=\frac{1}{\tau_i}e^{-t/\tau_i}, ~t>0,
\end{equation}
where $\psi_i(t)$ is the probability density function of dwell times $t>0$
and $\tau_i$ ($i=1,2$) are the mean dwell times in the two states (the
corresponding switching rates are then $1/\tau_i$). This case corresponds to
the state of the system alternating according to a Markov process, i.e.,
the switching between the two states is governed by a transition matrix.

\begin{figure}
\centering
\includegraphics[width=0.75\textwidth]{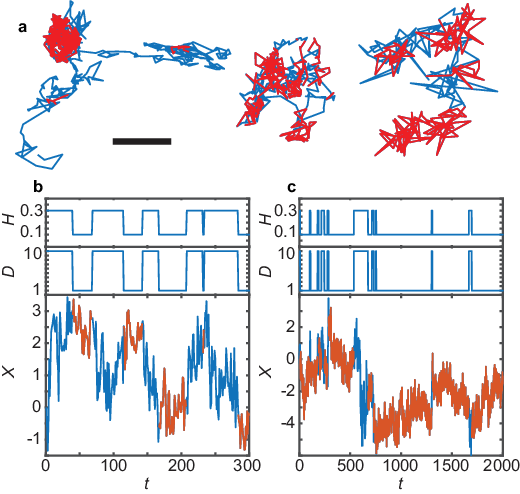}
\caption{\textbf{Representative switching FBM trajectories} \textcolor{blue}
{\textbf{a} Experimental trajectories obtained from single-particle tracking
of nanocrystals in live HeLa cells. The trajectories exhibit two states,
which are identified using the local convex hull, and shown in red and blue 
\cite{sabri2020elucidating}. The scale bar is 500 nm.} \textbf{b-c} Numerical 
simulations obtained using Eq.~\eqref{defFBMSE}. The two states in the trajectories are
$D_1=1$, $H_1=0.1$, and $D_2=10$, $H_2=0.3$. The upper panels show the Hurst
exponent and the diffusivity of the specific trajectories as a function of
time. \textcolor{blue}{Panel \textbf{b} shows} Markovian switching with $\tau=25$, 
\textcolor{blue}{while in panel \textbf{c}} one of the
states has a power-law waiting time distribution with $\beta=0.7$. The other
state has exponentially distributed waiting times with $\tau=15$. The time
in panel {\textbf b} is up to 300 in dimensionless units, while that in
panel {\textbf c} is up to 2,000, to emphasise the long dwell times in the
scale-free system.}
\label{Trajectories}
\end{figure}

We evaluate two different scenarios. In the first one, the Hurst exponent
$H$ remains constant and the generalised diffusion coefficient $D$ changes
according to a dichotomous Markov process, where the probability densities
of the dwell times are given by Eq.~\eqref{dwellMarkov}. In the second case,
also the Hurst exponent changes, thus, the two states are classified according
to their diffusivity $D_i$ and Hurst exponent $H_i$, where $i=1,2$ denotes
the state. It is futile to consider a special case where only $H$ changes
and $D$ remains constant because the units of $D$ depend on $H$, {\em vis},
$\mathrm{length}^2/\mathrm{time}^{2H}$. Therefore, even if one would attempt to
consider the same diffusivity in both states, they would still be different upon
a change of units such as transforming cm into $\mu$m. As a visual example
of the process, Fig.~\ref{Trajectories}b shows the first $300$ points of
a trajectory and the corresponding time series of $H$ and $D$.

A systematic evaluation of the two-state Markovian switching indicates that,
in the long time limit, the MSD is simply a weighted average of the MSDs
of the two original underlying processes. Given two states $D_i$ and $H_i$
with mean dwell times $\tau_i$, the MSD of the two parent FBM processes
are $\langle \overline{\delta_i^2(\Delta)} \rangle = 2D_i \Delta^{2H_i}$,
and the MSD of the switching FBM is
 \begin{equation}
 \label{msdMarkov}
 \left<\overline{\delta^2(\Delta)}\right>=A_1\left<\overline{\delta_1^2(\Delta)}
 \right>+A_2\left<\overline{\delta_2^2(\Delta)}\right>, 
 \end{equation}
where $A_i=\tau_i/(\tau_1+\tau_2)$.

\begin{figure}
\centering
\includegraphics[width=0.8\textwidth]{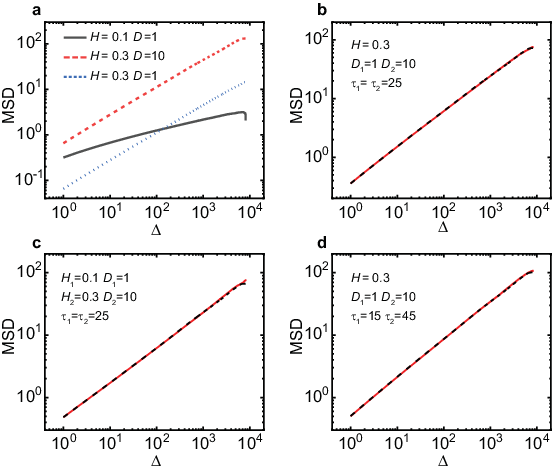}
\caption{\textbf{MSD of two-state FBM with Markovian switching exponents.}
{\textbf a} Standard (single-state) FBM simulated using Eq.~\eqref{defFBMSE}.
 \textbf{b} Two-state switching FBM where $D_1=1$ and $D_2=10$. The
 Hurst exponent is $H=0.3$ in both states and their mean dwell times are
 $\tau=25$. The dashed line indicates the average of the
 two underlying FBM processes.  \textbf{c} Two-state switching FBM where
 the parameters of state 1 are $D_1=1$, $H_1=0.1$, and the parameters of
 state 2 are $D_2=10$, $H_2=0.3$. The mean dwell times are $\tau=25$ for both
 states. The dashed line indicates the average of the two underlying
 FBM processes.  \textbf{d} Two-state switching FBM with the same parameters
 as in \textbf{b}, but with mean dwell times 15 and 45 in states 1 and
 2, respectively. The dashed line indicates the weighted average of the
 two underlying FBM processes, i.e., $\langle \overline{\delta^2(\Delta)}
 \rangle = 0.25 \langle \overline{\delta_1^2(\Delta)} \rangle +0.75 \langle
\overline{\delta_2^2(\Delta)} \rangle$.
}
\label{fig:MSD_markov}
\end{figure}

Fig.~\ref{fig:MSD_markov} shows the MSD of different simulations built from
states with $H=0.1, 0.3$ and $D=1, 10$. The MSD of the parent FBMs, i.e.,
without any switching, are shown in Fig.~\ref{fig:MSD_markov}a. Next,
Fig.~\ref{fig:MSD_markov}b shows the MSD when $D$ changes but
$H=0.3$ is kept constant and both states have the same mean dwell time
$\tau=25$. Interestingly, in this case, the anomalous diffusion exponent is
the same as that of the parent FBMs, $\alpha=2H$. Fig.~\ref{fig:MSD_markov}c
shows a case in which also $H$ changes, while the mean dwell times $\tau_i$
are the same in both states. In Fig.~\ref{fig:MSD_markov}d, the dwell
times are different, with $\tau_2 = 3\tau_1$. In all examined cases,
the MSD shows excellent agreement with the weighted average as given by
Eq.~\eqref{msdMarkov}.

\begin{figure}
\centering
\includegraphics[width=0.8\textwidth]{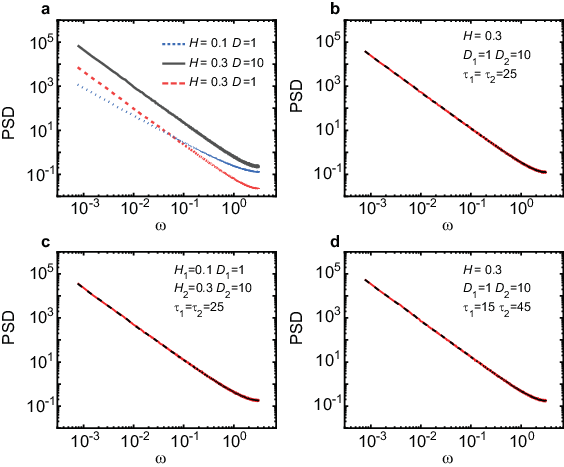}
\caption{\textbf{PSD of two-state FBM with Markovian switching exponents.}
{\textbf a} Standard (single-state) FBM simulated using Eq. \eqref{defFBMSE}.
\textbf{b} Two-state switching FBM where $D_1=1$ and $D_2=10$. The
Hurst exponent is $H=0.3$ in both states and their mean times are $\tau=25$. 
The dashed line indicates the average of the two underlying
FBM processes.  \textbf{c} Two-state switching FBM where the parameters of
state 1 are $D_1=1$, $H_1=0.1$, and the parameters of state 2 are $D_2=10$,
$H_2=0.3$. The mean times are $\tau=25$ for both states. The dashed line
indicates the average of the two underlying FBM processes.  \textbf{d}
Two-state switching FBM with the same parameters as in \textbf{b}, but with
time scales 15 and 45 for transitions from states 1 and 2, respectively. The
dashed line indicates the weighted average of the two underlying FBM processes,
i.e., $\langle S(\omega) \rangle = 0.25 \langle S_1(\omega) \rangle +0.75
\langle S_2(\omega) \rangle$.}
\label{fig:PSD_markov}
\end{figure}

The PSD of the switching FBM for two states having exponentially distributed
dwell times is shown in Fig.~\ref{fig:PSD_markov}. Following the same
structure as the MSD in Fig.~\ref{fig:MSD_markov}, the PSD of the parent
FBMs with $H=0.1, 0.3$ and $D=1, 10$ are shown in Fig.~\ref{fig:PSD_markov}a
and the PSD of the switching FBM alternating between these states are shown
in Figs.~\ref{fig:PSD_markov}b-d. These states correspond to subdiffusive
FBM. The PSD of FBM with $H>1/2$ depends on the observation time $T$
\cite{krapf2019spectral} and such cases for which the parent FBMs are
superdiffusive will be discussed later. Again, the PSD of the switching
process is given by the weighted average
\begin{equation}
\label{psdMarkov}
\langle S(\omega) \rangle = A_1 \langle S_1(\omega) \rangle +A_2 \langle S_2(\omega) \rangle, 
\end{equation}
where, once more, $A_i=\tau_i / (\tau_1 + \tau_2)$. The individual PSD of the
original subdiffusive FBM is $\langle S(\omega) \rangle \sim 1/\omega^{1+2H}$
and, thus, the switching FBM exhibits a similar spectral dependence,
\begin{equation}
\label{psdSFBM}
\langle S(\omega) \rangle \sim 1/\omega^{1+\alpha}.
\end{equation}

\subsection{Processes with scale-free relaxation times}

We now turn to study two-state dichotomous processes in which the dwell times
in one of the states are random variables with a heavy-tailed distribution,
namely, they are distributed according to a Pareto PDF,
\begin{equation}
\label{dwellPowerLaw}
\psi(t) = \frac{\beta t_0^\beta}{t^{1+\beta}}, \;\;\; t>t_0 , 
\end{equation}
with scale parameter $t_0>0$ and shape parameter $0<\beta<1$. The second
state is considered to have exponentially distributed dwell times. 
\textcolor{blue}{Such dichotomous processes, in which one of the states exhibits 
a dwell time distribution with an exponential tail and the second state has 
a power-law distribution, have received attention in diverse physical systems
\cite{sadegh20141,sikora2017elucidating,kurilovich2020complex,kurilovich2022non}.}
The first 2,000 points of a representative trajectory and its corresponding $H$
and $D$ time series are shown in Fig.~\ref{Trajectories}c.

\begin{figure}
\centering
\includegraphics[width=0.8\textwidth]{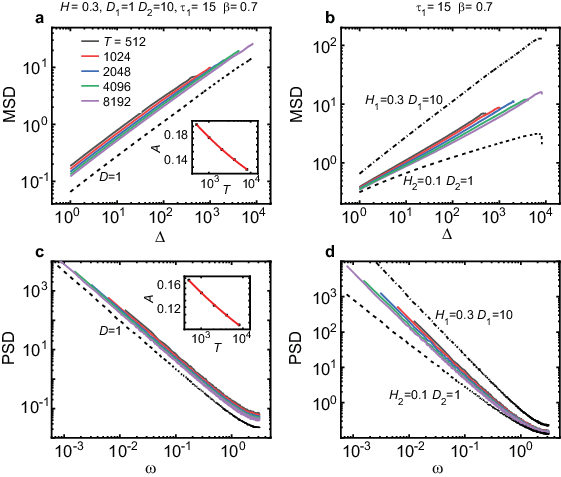}
\caption{\textbf{Two-state switching subdiffusive FBM for which one of the
states exhibits scale-free dwell times.} The MSD and PSD are computed for
different observation times $T$. As guides to the eye, the dashed lines
show the analysis of ordinary FBM processes. {\textbf a} MSD of two-state
switching FBM with the same Hurst exponent $H=0.3$ but different generalised
diffusion coefficients, such that $D_2=10 D_1$. The dwell times in the
first state are distributed according to a Pareto distribution, such that it
is asymptotically a power-law with exponent $\beta=0.7$. The second state
has an exponentially distributed dwell time with a mean $\tau=15$. The MSD
depends on the observation time. Inset: The amplitude of the MSD decays
as a power-law toward the MSD of the state with power-law waiting times, in
agreement with Eq.~\eqref{MSDpl1}. \textbf{b} Two-state FBM for which both
the Hurst exponent and the diffusivity change. Namely, $D_1=1$, $H_1=0.1$,
and the parameters of state 2 are $D_2=10$, $H_2=0.3$. The waiting times
in the first state have a heavy-tailed distribution. The second state has
exponentially distributed waiting times.  {\textbf c} PSD of the simulations
in panel {\textbf a}.  The inset shows that the amplitude of the PSD decays
as a power-law toward the PSD of the state with power-law waiting times,
in agreement with Eqs.~\eqref{MSDpl2} and \eqref{ampMSDpl2}. \textbf{d}
PSD of the simulations in panel {\textbf b}.}
\label{fig:MSD_power}
\end{figure}

Because one of the states has a dwell time with infinite mean,
the process is expected to exhibit ageing and ergodicity breaking
\cite{metzler2014anomalous,weron2017ergodicity,krapf2019spectral}.
Fig.~\ref{fig:MSD_power} shows the MSD and PSD of processes of this
type, for which the Hurst exponents of both states are subdiffusive,
$H_i<1/2$. The dependence on observation time is evident for both the MSD
and the PSD. Figs.~\ref{fig:MSD_power}a and c show, respectively, the MSD
and PSD of a system in which the Hurst exponent is the same for both states,
$H=0.3$, and the generalised diffusion coefficient changes 10-fold. The
obtained statistics yield
\begin{equation}
\left<\overline{\delta^2(\Delta,T)}\right>\sim(AT^{\beta-1}+2D_1)\Delta^{2H} 
\label{MSDpl1}
\end{equation}
and
\begin{equation}
\langle S(\omega,T) \rangle \sim \frac{C T^{\beta-1} +S_1}{\omega^{1+2H}},
\label{PSDpl1}
\end{equation}
where state 1, is the one with power-law sojourn times. The amplitude of
the MSD (PSD) of the switching FBM is such that it slowly approaches (in a
power-law) to the amplitude of the MSD (PSD) of state 1, see the insets of
Figs.~\ref{fig:MSD_power}a and c. To be precise, the MSD converges to $2D_1
\Delta^{2H}$ and the PSD to $S_1/\omega^{1+2H}$, where $S_1=2D_1\Gamma(2H +
1) \sin(\pi H)$ \cite{krapf2019spectral}. For any experimental time $T$, in the
long lag-time limit, the MSD scales as $\Delta^{2H}$ and the PSD scales
as $\omega^{-(1+2H)}$.

When the Hurst exponents of the two states are different, the MSD and PSD still
converge towards those of the state with power-law sojourn times. However,
the results are fairly different in that, now, the MSD dependence on lag
time $\Delta$ and the frequency dependence of the PSD have exponents that
depend on the experimental time $T$. In this case,
\begin{equation}
\left<\overline{\delta^2(\Delta,T)}\right>\sim A(T)\Delta^{\alpha(T)} 
\label{MSDpl2}
\end{equation}
and
\begin{equation}
\langle S(\omega,T) \rangle \sim \frac{C(T)}{\omega^{1+\alpha(T)}},
\label{PSDpl2}
\end{equation}
where the amplitudes $A(T)$ and $C(T)$, and the exponents $\alpha(T)$ are given
by
\begin{align}
A(T) &= A_0 T^{\beta-1} + 2D_1, \nonumber \\
C(T) &= C_0 T^{\beta-1} + S_1,
\label{ampMSDpl2}
\end{align}
and
\begin{equation}
\alpha(T)=\alpha_0 T^{\beta-1} +\alpha_1,
\label{alphaPL2}
\end{equation}
where $A_0$, $C_0$, and $\alpha_0$ are constants that depend on the occupation
fraction in state 1 during the initial time of the process and $\alpha_1 =
2H_1$ is the anomalous diffusion exponent of state 1.

\subsection{Switching superdiffusive FBM}

FBM can be subdiffusive ($H<1/2$) or superdiffusive ($H>1/2$). While the
MSD in both cases scales as $\langle \overline{\delta^2(\Delta)} \rangle
\sim \Delta^{2H}$, the scaling of the PSD differs among the two classes. As
discussed above, for subdiffusive FBM, $\langle S(\omega) \rangle \sim
1/\omega^{1+2H}$, but when the FBM is superdiffusive, the frequency scaling
of the PSD resembles that of Brownian motion, albeit with a dependence
on observation time, $\langle S(\omega,T) \rangle \sim T^{2H-1}/\omega^2$
\cite{krapf2019spectral}.  Therefore, the analysis of switching superdiffusive
FBM needs separate attention.

\begin{figure}
\centering
\includegraphics[width=0.8\textwidth]{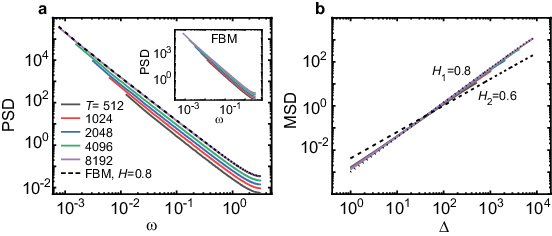}
\caption{\textbf{Two-state superdiffusive FBM for which one of the states
exhibits scale-free dwell times.} The PSD (panel \textbf{a}) and MSD (panel
\textbf{b}) are computed for different realisation times. The process
consists of two states with $H_1=0.6$ and $H_2=0.8$. The sojourn times in
state 1 are exponentially distributed and those of state 2 have a heavy-tailed
distribution with $\beta=0.7$. The inset in panel \textbf{a} shows the ageing
in the traditional FBM for $H=0.8$ in dependence of the observation time $T$.}
\label{fig:super}
\end{figure}

Fig.~\ref{fig:super} shows the PSD and MSD of switching FBM  consisting of
two states with the Hurst exponents $H_1=0.6$ and $H_2=0.8$, and power-law
distributed sojourn times in state 2. The outcome involves a dependence on
observation time that arises  from both the switching mechanism and the FBM
itself. Namely, the PSD has a dependence on frequency of the form $1/\omega^2$,
and a dependence on experimental time with a scaling factor $T^{\alpha(T)-1}$
from the FBM and a scaling factor $(C T^{\beta-1} +S_1)$ due to the
switching between states. In addition, the MSD involves a change in the
anomalous diffusion exponent $\alpha$, similar to that in Eq.~\eqref{alphaPL2}.
\textcolor{blue}{When the first state is subdiffusive ($H_1<0.5$,
data not shown) the obtained results do not exhibit any difference from
those when both processes are superdiffusive. Namely for the mixed sub-
and superdiffusive case, the PSD and MSD asymptotically converge to those
of the state with power-law dwell times.}

\subsection{Analysis of experimental data: quantum dots in the cytoplasm of
live cells}

In order to highlight the use of the switching FBM process in the analysis
of real world data, we analyse the PSD of quantum dot trajectories in
the cytoplasm of living HeLa cells. These data have been thoroughly
analysed in terms of their MSD, velocity autocorrelation function, and
distribution of displacements \cite{sabri2020elucidating}, as well as via
the use of a hidden Markov model approach \cite{janczura2021identifying},
the intermediate scattering function \cite{dieball2022scattering}
and the decomposition of the Hurst exponent into components involving
non-stationarity, heavy-tailed distributions, and long-range correlations
\cite{vilk2022unravelling}. These extensive analyses show that the diffusive
motion of quantum dots stochastically alternates between two states, with
both states having correlations of the type of subdiffusive FBM. Thus,
quantum dot dynamics in the cytoplasm presents an excellent system to test
some of the predictions of the switching FBM model. The switching between
the two states in this experimental system obeys a Markov process and the
MSD is subdiffusive with a mean anomalous diffusion exponent $\alpha=0.59$
\cite{sabri2020elucidating,janczura2021identifying}. Our predictions indicate
that the PSD should not exhibit ageing effects and its spectral dependence,
according to Eq.~\eqref{psdSFBM}, is expected to be $\langle S(\omega)
\rangle \sim 1/\omega^{1+\alpha}$.

The difficulty in the analysis of experimental data lies in the fact that
long trajectories are not available because eventually particles leave the
field of view, or they become dark (due to long blinking in the case of
quantum dots or photobleaching in the case of organic fluorophores). The
analysed quantum dot data consist of 3,834 trajectories of only 100
time points each. Such short trajectories present unique problems in the
statistical analysis. Further, experimental data is unavoidably corrupted by
experimental noise, such as static and dynamic localisation errors inherent
to single-particle tracking \cite{savin2005static}.

\begin{figure}
\centering
\includegraphics[width=0.8\textwidth]{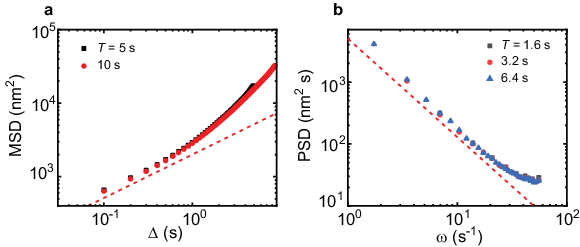}
\caption{\textbf{Analysis of experimental trajectories of quantum dots
in the cytoplasm of live mammalian cells.} The projection of quantum dot
trajectories along the $x$ axis is analysed, for different measurement times $T$,
in terms of \textbf{a} the MSD and
\textbf{b} the PSD. For different measurement times, the MSDs overlap (no ageing)
and exhibits a behaviour $\langle \overline{\delta^2(\Delta)} \rangle \sim
\Delta^{0.59}$ at short lag-times, indicating an apparent anomalous diffusion
exponent $\alpha=0.59$, as obtained previously by averaging the anomalous
diffusion exponents of individual trajectories \cite{sabri2020elucidating}. In
agreement with the predictions for Markov switching, the PSDs overlap for
the three measurement times and show a scaling $\langle S(\omega,T) \rangle
\sim 1/\omega^{1.59}$. The dashed red lines are guides to the eye indicating
a scaling $\Delta^{0.59}$ in panel \textbf{a} and $\omega^{1.59}$ in panel \textbf{b}.}
\label{fig:QD}
\end{figure}

The MSD and PSD analysis of quantum dot trajectories along the projections on
one axis is presented in Fig.~\ref{fig:QD}. The MSD at short times is seen to
scale as $\langle \overline{\delta^2(\Delta)} \rangle \sim \Delta^{\alpha}$
with $\alpha=0.59$. Despite the short length of the trajectories and the
presence of localisation errors, the agreement with the predicted PSD is
remarkable. The analysis is performed for three observation times ($T=1.6$
s, $3.2$ s, and $6.4$ s) consisting of 16, 32, and 64 time points. The
three PSDs are observed to fall on the same line, i.e., there is no evident
ageing, and the slope of the PSD agrees with the prediction $1/\omega^{1.59}$.

\subsection{Distribution of displacements}

\begin{figure}
\centering
\includegraphics[width=0.8\textwidth]{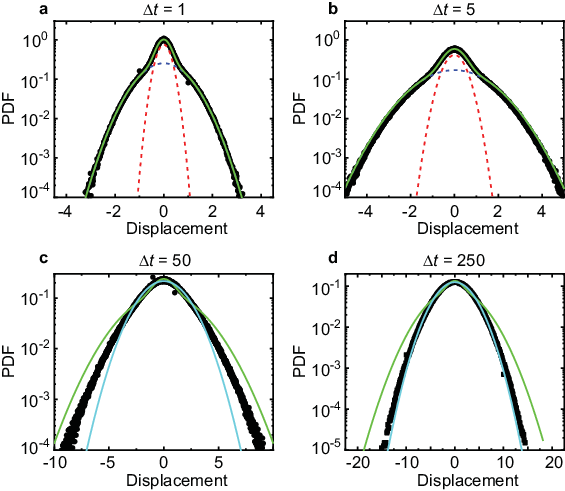}
\caption{\textbf{Distributions of displacements for switching FBM process
are not Gaussian.} The PDF of the displacements is estimated for a process
with Markov switching, whose mean dwell time in each state is $\tau=25$, the
Hurst exponent of both states is $H=0.3$ and the diffusivities are $D_1=1$
and $D_2=10$. The times over which the displacements are computed are {\textbf
(a)} $\Delta t=1$, {\textbf (b)} $\Delta t=5$, {\textbf (c)} $\Delta t=50$,
and  {\textbf (d)} $\Delta t=250$. In panels {\textbf (a)} and {\textbf (b)}
(shoter times), the two coloured dashed lines show a decomposition into
two Gaussian functions with the standard deviations of the distributions
of the two parent FBM. The green solid line indicates the superposition
of the two Gaussian functions. In panels {\textbf (c)} and {\textbf (d)}
(longer times), the solid green line shows a superposition of two Gaussian
functions and the thick cyan line shows a fit to a single Gaussian peak.}
\label{fig:Displacements}
\end{figure}

Both the MSD and PSD of switching FBM with exponentially-distributed dwell
times resemble those of classical FBM, making it impossible to rely solely on
these statistics to identify the model. The problem is less severe when the
distribution of dwell times have heavy tails because in these cases, the MSD
and PSD exhibit ageing in stark contrast to ordinary FBM processes. One clear
signature of heterogeneous or intermittent processes lies in the distribution
of displacements which is typically non-Gaussian. Fig.~\ref{fig:Displacements}
shows the distribution of displacements for switching FBM realisations
with exponential dwell time distribution. Here we present a process with
characteristic dwell times $\tau=25$ and displacements over times that span
a scale from much shorter to much longer times than this characteristic
time. Namely, displacements were computed at four different times, $\Delta
t=1$, $5$, $50$, and $250$.

For times much shorter than the characteristic time
(Figs.~\ref{fig:Displacements}a and b, $\Delta t\ll\tau$), the distribution of
displacements is very close to the sum of two Gaussian functions. Specifically,
the two parent FBM have a normal distribution of displacements with standard
deviations $\sigma_1$ and $\sigma_2$. Then, the switching FBM process has a
distribution that is a sum of two Gaussians with the same standard deviations,
namely $P_{\Delta t}(\Delta x)=\sum_i A_i \exp(\Delta x^2/2\sigma_i^2)$,
where $\sigma_i^2\propto\Delta t^{2H_i}$.

For times longer than the characteristic dwell time ($\Delta t >\tau$),
the situation is rather different. In fact, as the times over which the
displacements are computed become much longer than the characteristic time
($\Delta t \gg \tau$), the distribution of displacements approaches a normal
distribution. Fig.~\ref{fig:Displacements}d shows the behaviour at time
$\Delta t =250$, i.e., $\Delta t = 10 \tau$, and here the deviations from
Gaussianity are very small.

\section{Discussion and Conclusions}

We studied FBM, a stochastic process driven by long-ranged correlated
Gaussian noise, in which both diffusion coefficient and Hurst exponent are
stochastic processes themselves. We modelled these for cases with exponential
and scale-free dwell time distributions. This model belongs to the class
of doubly-stochastic processes (both the driving noise but also the model
parameters are stochastic) that currently receive increased attention. In
particular, we analysed the time-averaged MSD and the PSD of the emerging
dynamics.

Markovian switching resembles ordinary FBM both in terms of the PSD and
the MSD. Thus, these metrics are not sufficient to dissect the process
and recognise that it is not driven by a single FBM. In other words, the
switching dynamic does not have any clear fingerprint in the MSD or the
PSD. An additional statistic that provides information on the switching
can be obtained from the distribution of displacements, which in this
case is non-Gaussian (Fig.~\ref{fig:Displacements}). Such distributions have also
been observed experimentally, e.g., in quantum dot trajectories in the
cytoplasm for which the non-Gaussian nature of the displacement hints at a
more complex process than FBM. However, as the displacements are computed
over increasingly longer times in the switching FBM model, deviations from
Gaussianity subside. Such effects are also observed for experimental data,
where, as time increases, the distribution of displacements approaches a Gaussian
distribution \cite{sabri2020elucidating}. Thus, in order to identify a Markov
switching process, it is necessary to obtain measurements with a temporal
resolution better than the characteristic dwell times. If this type of data is
available, the two states can be identified using a change point detection tool
\cite{lanoiselee2017unraveling,sikora2017elucidating,wagner2017classification}.

Going beyond Markov switching, scale-free processes, in which at least one of
the states has a heavy-tailed distribution of dwell times are inherently
non-ergodic and have non-stationary increments. The quintessential process
of this type is the continuous time random walk \cite{metzler2014anomalous}.
When one of the states has a heavy-tailed distribution of dwell times, both
the MSD and the PSD depend explicitly on time.

An interesting case is that of superdiffusive FBM, 
\textcolor{blue}{based on persistent active stochastic dynamics, such as intracellular
motion driven by molecular motors in living cells \cite{reverey2015superdiffusion} or animal motion
\cite{vilk2022unravelling,vilk2022ergodicity}. This should not be taken to indicate that active motion will always
lead to superdiffusion; as soon as there exists a finite persistence time,
the motion will be Brownian at times longer than this time scale \cite{romanczuk2012active,lemaitre2023non}.
Moreover, transient superdiffusion may arise in passive systems, such as
bulk-mediated diffusion \cite{campagnola2015superdiffusive}.}
Superdiffusive FBM has a PSD that
depends on experimental time $T$ \cite{krapf2019spectral} and, thus, also the
Markovian switching for which at least one of the states is superdiffusive
exhibits a PSD that depends on realisation time. However, the PSD is still
found as a weighted average of the parent (time-dependent) PSDs.

A simple visual inspection allows one to determine whether the two states
in the switching FBM have the same Hurst exponent $H$. In the case that $H$
does not change, the anomalous diffusion exponent as determined by the PSD
and the MSD does not depend on the experimental time $T$. In such cases,
the MSD (and PSD) exhibit the same slope when visualised in a log-log plot
for different realisation times (Fig.~\ref{fig:MSD_power}a). The MSD (PSD)
converges to the MSD (PSD) of the FBM with power-law waiting times. However,
the convergence has a power-law character. When $H$ changes, the MSD (PSD)
still converges to the state with heavy-tailed waiting times but in this
case each experimental time exhibits a different exponent.
\textcolor{blue}{We foresee that switching FBM with different Hurst exponents 
can have multiple direct applications in cell biology, 
such as the heterogeneous dynamics of intracellular endosomes \cite{fedotov}.} 

To simplify the analyses, we restricted our work in switching FBM to two
states. However, there is no actual limit to the number of states that
can be included. In particular, a multi-state Markov process can include a
full transition matrix between the different states. This work opens the
way to modelling heterogeneous anomalous dynamics, where the underlying
heterogeneity leads to dynamic transitions. Moreover, the results obtained
allow for future theoretical investigations of correlated random walks in
complex systems where regime changes dominate the transport.

\begin{acknowledgments}

The experimental data were obtained in collaboration with Matthias Weiss,
Adal Sabri, and Xinran Xu. D.K. thanks O'Neil Wiggan for providing the HeLa
cells. D.K. acknowledges funding from the National Science Foundation grant
2102832. R.M. acknowledges funding from German Science Foundation (DFG,
grant ME 1535/12-1). A.W. acknowledges  National Center of Science (Poland)
- Opus Grant 2020/37/B/HS4/00120.

\end{acknowledgments}

\section*{References}

\providecommand{\newblock}{}

\section*{Appendix A. Extended numerical methods}

Here, we describe the simulation procedure of the switching FBM.

\subsection{General case.}

We first consider the general case since Eq.~\eqref{defFBMSE}, which defines switching FBM, is a special case of the integral representation of the process
\begin{align}
\label{eq:int_representation}
X(t)=\int_{-\infty}^\infty f_t(s)\mathrm{d}B(s).
\end{align}
To simulate the process $X(t)$, we need to numerically approximate the integral \eqref{eq:int_representation}. Usually, this is done in two steps:
first, truncating the limits of integration (this step depends on the form of $f_t$),
and, second, approximating the truncated integral by a Riemann sum.

\noindent First step. For $M_1, M_2 \in \mathbb{R}$ we have
\begin{align}
\int_{-\infty}^\infty f_t(s) \mathrm{d}B(s) \approx \int_{M_1}^{M_2} f_t(s) \mathrm{d}B(s).
\end{align}
In general, it is advised to choose reasonably large values $M_1$ and $M_2$. In practice, the truncation parameters may also depend on $t$.\\
Second step. We divide the interval $[M_1, M_2]$ into $I$ equal parts of length $\lambda = \frac{M_2-M_1}{I}$, and consider points $s_i = M_1 + i\lambda$ for $i=0,1,\ldots, I$. Then
\begin{align}
\int_{M_1}^{M_2} f_t(s) \mathrm{d}B(s) = \sum_{i=0}^{I-1} \int_{s_i}^{s_{i+1}} f_t(s) \mathrm{d}B(s).
\end{align}
Now, we assume that each subinterval $\left[s_i, s_{i+1}\right)$ is small ($I$ is large) and we apply a Riemann (or Euler) type of approximation, i.e., we calculate the function $f_t(s)$ at $s_i$ for $i=0, 1, \ldots,$ $ I-1$ and obtain
\begin{align}
\label{eq:int_approximation}
X(t) \approx \sum_{i=0}^{I-1}\int_{s_i}^{s_{i+1}} f_t(s_i) \mathrm{d}B(s) = \sum_{i=0}^{I-1} f_t(s_i)\left[B(s_{i+1}) - B(s_i)\right]=\sum_{i=0}^{I-1} f_t(s_i)\xi_i,
\end{align}
where $\xi_i$ are i.i.d. random variables with $N(0,\lambda)$ distribution.

Finally, we note that to simulate a trajectory of the process $X(t)$ at various time points $t_1, t_2, \ldots, t_n$, we rely on one sequence of $\xi_i$' s (a single trajectory of the process $B(s)$ for all possible values of $s$).

\subsection{Switching FBM.} In this particular case, to generate a single
trajectory of switching FBM we first need to generate trajectories of the processes $D(s)$ and $H(s)$. We do this by generating waiting times from the selected distribution. Then,
having a single trajectory of both $D$ and $H$ we use the approximation
given in Eq.~\eqref{eq:int_approximation} with
\begin{align*}
f_t(s)=\sqrt{2D(s)H(s)}(t-s)^{H(s)-\frac{1}{2}}\mathbf{1}_{\left[0,t\right)}(s),\;t\in [0,T].
\end{align*}
The points $s_0, s_1, \ldots, s_I$ are chosen so that $s_0 = M_1 = 0$ and $s_I = M_2 = T$. In all simulations performed, the length of the trajectories $n = 2^{13}$, $t_i = i \Delta_t, i = 0, 1, \ldots, n$, $\Delta_t =
\frac{T}{n}$ and $I = 50 n$. Thus, to calculate a next time step $X(t_j)$ we use an additional 50 new steps $s_i$, e.g., to calculate $X(t_1)$ we use $f_{t_1}(s)$ in points $s_0, s_1, \ldots, s_{49}$, to calculate $X(t_2)$ we use $f_{t_2}(s)$ in points $s_0, s_1, \ldots, s_{49}, s_{50}, \ldots, s_{99}$, and so on.



\begin{thebibliography}{10}
\expandafter\ifx\csname url\endcsname\relax
  \def\url#1{{\tt #1}}\fi
\expandafter\ifx\csname urlprefix\endcsname\relax\def\urlprefix{URL }\fi
\providecommand{\eprint}[2][]{\url{#2}}

\bibitem{levi2007exploring}
Levi V and Gratton E 2007 {\em Cell Biochemistry and Biophysics\/} {\bf 48}
  1--15

\bibitem{manzo2015review}
Manzo C and Garcia-Parajo M~F 2015 {\em Reports on Progress in Physics\/} {\bf
  78} 124601

\bibitem{barkai2012strange}
Barkai E, Garini Y and Metzler R 2012 {\em Physics Today\/} {\bf 65(8)} 29

\bibitem{hofling2013anomalous}
H{\"o}fling F and Franosch T 2013 {\em Reports on Progress in Physics\/} {\bf
  76} 046602

\bibitem{krapf2019strange}
Krapf D and Metzler R 2019 {\em Physics Today\/} {\bf 72(9)} 48--54

\bibitem{dietrich2002relationship}
Dietrich C, Yang B, Fujiwara T, Kusumi A and Jacobson K 2002 {\em Biophysical
  Journal\/} {\bf 82} 274--284

\bibitem{knight2009single}
Knight J~D and Falke J~J 2009 {\em Biophysical Journal\/} {\bf 96} 566--582

\bibitem{campagnola2015superdiffusive}
Campagnola G, Nepal K, Schroder B~W, Peersen O~B and Krapf D 2015 {\em
  Scientific Reports\/} {\bf 5} 17721

\bibitem{manzo2015weak}
Manzo C, Torreno-Pina J~A, Massignan P, Lapeyre~Jr G~J, Lewenstein M and
  Garcia~Parajo M~F 2015 {\em Physical Review X\/} {\bf 5} 011021

\bibitem{metz2019temporal}
Metz M~J, Pennock R~L, Krapf D and Hentges S~T 2019 {\em Scientific Reports\/}
  {\bf 9} 7297

\bibitem{mosqueira2020antibody}
Mosqueira A, Camino P~A and Barrantes F~J 2020 {\em Journal of
  Neurochemistry\/} {\bf 152} 663--674

\bibitem{weigel2011ergodic}
Weigel A~V, Simon B, Tamkun M~M and Krapf D 2011 {\em Proceedings of the
  National Academy of Sciences\/} {\bf 108} 6438--6443

\bibitem{akin2016single}
Akin E~J, Sol{\'e} L, Johnson B, El~Beheiry M, Masson J~B, Krapf D and Tamkun
  M~M 2016 {\em Biophysical Journal\/} {\bf 111} 1235--1247

\bibitem{he2016dynamic}
He W, Song H, Su Y, Geng L, Ackerson B~J, Peng H and Tong P 2016 {\em Nature
  Communications\/} {\bf 7} 11701

\bibitem{bronstein2009transient}
Bronstein I, Israel Y, Kepten E, Mai S, Shav-Tal Y, Barkai E and Garini Y 2009
  {\em Physical Review Letters\/} {\bf 103} 018102

\bibitem{moon2019multicolour}
Moon S~L, Morisaki T, Khong A, Lyon K, Parker R and Stasevich T~J 2019 {\em
  Nature Cell Biology\/} {\bf 21} 162--168

\bibitem{ruhnow2011tracking}
Ruhnow F, Zwicker D and Diez S 2011 {\em Biophysical Journal\/} {\bf 100}
  2820--2828

\bibitem{nixon2016increased}
Nixon-Abell J, Obara C~J, Weigel A~V, Li D, Legant W~R, Xu C~S, Pasolli H~A,
  Harvey K, Hess H~F, Betzig E {\em et~al.\/} 2016 {\em Science\/} {\bf 354}
  aaf3928

\bibitem{speckner2018anomalous}
Speckner K, Stadler L and Weiss M 2018 {\em Physical Review E\/} {\bf 98}
  012406

\bibitem{korabel2021local}
Korabel N, Han D, Taloni A, Pagnini G, Fedotov S, Allan V and Waigh T~A 2021
  {\em Entropy\/} {\bf 23} 958

\bibitem{weihs2006bio}
Weihs D, Mason T~G and Teitell M~A 2006 {\em Biophysical Journal\/} {\bf 91}
  4296--4305

\bibitem{etoc2018non}
Etoc F, Balloul E, Vicario C, Normanno D, Li{\ss}e D, Sittner A, Piehler J,
  Dahan M and Coppey M 2018 {\em Nature Materials\/} {\bf 17} 740--746

\bibitem{sabri2020elucidating}
Sabri A, Xu X, Krapf D and Weiss M 2020 {\em Physical Review Letters\/} {\bf
  125} 058101

\bibitem{berg2000motile}
Berg H~C 2000 {\em Physics Today\/} {\bf 53} 24--29

\bibitem{sims2019optimal}
Sims D~W, Humphries N~E, Hu N, Medan V and Berni J 2019 {\em Elife\/} {\bf 8}
  e50316

\bibitem{hays2012high}
Hays G~C, Bastian T, Doyle T~K, Fossette S, Gleiss A~C, Gravenor M~B, Hobson
  V~J, Humphries N~E, Lilley M~K, Pade N~G {\em et~al.\/} 2012 {\em Proceedings
  of the Royal Society B: Biological Sciences\/} {\bf 279} 465--473

\bibitem{vilk2022unravelling}
Vilk O, Aghion E, Avgar T, Beta C, Nagel O, Sabri A, Sarfati R, Schwartz D~K,
  Weiss M, Krapf D {\em et~al.\/} 2022 {\em Physical Review Research\/} {\bf 4}
  033055

\bibitem{vilk2022ergodicity}
Vilk O, Orchan Y, Charter M, Ganot N, Toledo S, Nathan R and Assaf M 2022 {\em
  Physical Review X\/} {\bf 12} 031005

\bibitem{bouchaud2005subtle}
Bouchaud J~P 2005 {\em Chaos: An Interdisciplinary Journal of Nonlinear
  Science\/} {\bf 15} 026104

\bibitem{scalas2006application}
Scalas E 2006 {\em Physica A: Statistical Mechanics and its Applications\/}
  {\bf 362} 225--239

\bibitem{edery2010particle}
Edery Y, Scher H and Berkowitz B 2010 {\em Water Resources Research\/} {\bf 46}
  W07524

\bibitem{weigel2012obstructed}
Weigel A~V, Ragi S, Reid M~L, Chong E~K, Tamkun M~M and Krapf D 2012 {\em
  Physical Review E\/} {\bf 85} 041924

\bibitem{wu2020nanoparticle}
Wu H and Schwartz D~K 2020 {\em Accounts of Chemical Research\/} {\bf 53}
  2130--2139

\bibitem{metzler2014anomalous}
Metzler R, Jeon J~H, Cherstvy A~G and Barkai E 2014 {\em Physical Chemistry
  Chemical Physics\/} {\bf 16} 24128--24164

\bibitem{krapf2015mechanisms}
Krapf D 2015 {\em Current Topics in Membranes\/} {\bf 75} 167--207

\bibitem{krapf2018power}
Krapf D, Marinari E, Metzler R, Oshanin G, Xu X and Squarcini A 2018 {\em New
  Journal of Physics\/} {\bf 20} 023029

\bibitem{yazmin}
Meroz Y and Sokolov I~M 2015 {\em Physics Reports\/} {\bf 573} 1--29

\bibitem{burrage}
Robson A, Burrage K and Leake M~C 2013 {\em Philosophical Transactions of the
  Royal Society B: Biological Sciences\/} {\bf 368} 20120029

\bibitem{thapa}
Thapa S, Lomholt M~A, Krog J, Cherstvy A~G and Metzler R 2018 {\em Physical
  Chemistry Chemical Physics\/} {\bf 20} 29018--29037

\bibitem{munoz2021objective}
Mu{\~n}oz-Gil G, Volpe G, Garcia-March M~A, Aghion E, Argun A, Hong C~B, Bland
  T, Bo S, Conejero J~A, Firbas N {\em et~al.\/} 2021 {\em Nature
  Communications\/} {\bf 12} 6253

\bibitem{henrik}
Seckler H and Metzler R 2022 {\em Nature Communications\/} {\bf 13} 6717

\bibitem{henrik1}
Seckler H, Szwabi{\'n}ski J and Metzler R 2023 {\em Journal of Physical
  Chemistry Letters\/} {\bf 14} 7910

\bibitem{janusz}
Gajowczyk M and Szwabi{\'n}ski J 2021 {\em Entropy\/} {\bf 23} 649

\bibitem{gorka}
Mu{\~n}oz-Gil G, Garcia-March M~A, Manzo C, Mart{\'\i}n-Guerrero J~D and
  Lewenstein M 2020 {\em New Journal of Physics\/} {\bf 22} 013010

\bibitem{montroll1965random}
Montroll E~W and Weiss G~H 1965 {\em Journal of Mathematical Physics\/} {\bf 6}
  167--181

\bibitem{scher1973stochastic}
Scher H and Lax M 1973 {\em Physical Review B\/} {\bf 7} 4491

\bibitem{scher1991time}
Scher H, Shlesinger M~F and Bendler J~T 1991 {\em Physics Today\/} {\bf 44}
  26--34

\bibitem{kolmogorov1940wienersche}
Kolmogorov A~N 1940 {\em Acad. Sci. URSS (NS)\/} {\bf 26} 115--118

\bibitem{mandelbrot1968fractional}
Mandelbrot B~B and Van~Ness J~W 1968 {\em SIAM Review\/} {\bf 10} 422--437

\bibitem{szymanski2009elucidating}
Szymanski J and Weiss M 2009 {\em Physical Review Letters\/} {\bf 103} 038102

\bibitem{magdziarz2009fractional}
Magdziarz M, Weron A, Burnecki K and Klafter J 2009 {\em Physical Review
  Letters\/} {\bf 103} 180602

\bibitem{sadegh2017plasma}
Sadegh S, Higgins J~L, Mannion P~C, Tamkun M~M and Krapf D 2017 {\em Physical
  Review X\/} {\bf 7} 011031

\bibitem{lampo2017cytoplasmic}
Lampo T~J, Stylianidou S, Backlund M~P, Wiggins P~A and Spakowitz A~J 2017 {\em
  Biophysical Journal\/} {\bf 112} 532--542

\bibitem{jeon2016protein}
Jeon J~H, Javanainen M, Martinez-Seara H, Metzler R and Vattulainen I 2016 {\em
  Physical Review X\/} {\bf 6} 021006

\bibitem{Balcerek2023}
Balcerek M, Burnecki K, Thapa S, Wy{\l}oma{\'n}ska A and Chechkin A 2022 {\em
  Chaos: An Interdisciplinary Journal of Nonlinear Science\/} {\bf 32} 093114

\bibitem{jeon2011vivo}
Jeon J~H, Tejedor V, Burov S, Barkai E, Selhuber-Unkel C, Berg-S{\o}rensen K,
  Oddershede L and Metzler R 2011 {\em Physical review letters\/} {\bf 106}
  048103

\bibitem{tabei2013intracellular}
Tabei S~A, Burov S, Kim H~Y, Kuznetsov A, Huynh T, Jureller J, Philipson L~H,
  Dinner A~R and Scherer N~F 2013 {\em Proceedings of the National Academy of
  Sciences\/} {\bf 110} 4911--4916

\bibitem{deng2009ergodic}
Deng W and Barkai E 2009 {\em Physical Review E\/} {\bf 79} 011112

\bibitem{choquet2013dynamic}
Choquet D and Triller A 2013 {\em Neuron\/} {\bf 80} 691--703

\bibitem{weigel2013quantifying}
Weigel A~V, Tamkun M~M and Krapf D 2013 {\em Proceedings of the National
  Academy of Sciences\/} {\bf 110} E4591--E4600

\bibitem{sikora2017elucidating}
Sikora G, Wy{\l}oma{\'n}ska A, Gajda J, Sol{\'e} L, Akin E~J, Tamkun M~M and
  Krapf D 2017 {\em Physical Review E\/} {\bf 96} 062404

\bibitem{weron2017ergodicity}
Weron A, Burnecki K, Akin E~J, Sol{\'e} L, Balcerek M, Tamkun M~M and Krapf D
  2017 {\em Scientific Reports\/} {\bf 7} 5404

\bibitem{arcizet2008temporal}
Arcizet D, Meier B, Sackmann E, R{\"a}dler J~O and Heinrich D 2008 {\em
  Physical Review Letters\/} {\bf 101} 248103

\bibitem{fedotov}
Han D, Korabel N, Chen R, Johnston M, Gavrilova A, Allan V~J, Fedotov S and
  Waigh T~A 2020 {\em Elife\/} {\bf 9} e52224

\bibitem{loverdo2009quantifying}
Loverdo C, Benichou O, Voituriez R, Biebricher A, Bonnet I and Desbiolles P
  2009 {\em Physical Review Letters\/} {\bf 102} 188101

\bibitem{andreao2006ecg}
Andreao R~V, Dorizzi B and Boudy J 2006 {\em IEEE Transactions on Biomedical
  Engineering\/} {\bf 53} 1541--1549

\bibitem{khanagha2014phonetic}
Khanagha V, Daoudi K, Pont O and Yahia H 2014 {\em Digital Signal Processing\/}
  {\bf 35} 86--94

\bibitem{cetin2006short}
Cetin M and Comert G 2006 {\em Transportation Research Record\/} {\bf 1965}
  23--31

\bibitem{janczura2013goodness}
Janczura J and Weron R 2013 {\em AStA Advances in Statistical Analysis\/} {\bf
  97} 239--270

\bibitem{lux2010forecasting}
Lux T and Morales-Arias L 2010 {\em Computational Statistics \& Data
  Analysis\/} {\bf 54} 2676--2692

\bibitem{edelhoff2016path}
Edelhoff H, Signer J and Balkenhol N 2016 {\em Movement Ecology\/} {\bf 4}
  1--21

\bibitem{stanislavsky2009farima}
Stanislavsky A, Burnecki K, Magdziarz M, Weron A and Weron K 2009 {\em The
  Astrophysical Journal\/} {\bf 693} 1877

\bibitem{vasas2007two}
Vasas K, Elek P and M{\'a}rkus L 2007 {\em Journal of Statistical Planning and
  Inference\/} {\bf 137} 3113--3126

\bibitem{levy1953random}
L{\'e}vy P 1953 {\em Random functions: general theory with special reference to
  Laplacian random functions\/} 12 (University of California Press)

\bibitem{wang2023memory}
Wang W, Balcerek M, Burnecki K, Chechkin A~V, Janu\v{s}onis S, {\'S}l\k{e}zak
  J, Vojta T, Wy{\l}oma{\'n}ska A and Metzler R 2023 {\em Phys. Rev. Res.\/}
  {\bf 5} L032025

\bibitem{krapf2019spectral}
Krapf D, Lukat N, Marinari E, Metzler R, Oshanin G, Selhuber-Unkel C, Squarcini
  A, Stadler L, Weiss M and Xu X 2019 {\em Physical Review X\/} {\bf 9} 011019

\bibitem{sadegh20141}
Sadegh S, Barkai E and Krapf D 2014 {\em New Journal of Physics\/} {\bf 16}
  113054

\bibitem{kurilovich2020complex}
Kurilovich A~A, Mantsevich V~N, Stevenson K~J, Chechkin A~V and Palyulin V~V
  2020 {\em Physical Chemistry Chemical Physics\/} {\bf 22} 24686--24696

\bibitem{kurilovich2022non}
Kurilovich A~A, Mantsevich V~N, Mardoukhi Y, Stevenson K~J, Chechkin A~V and
  Palyulin V~V 2022 {\em Physical Chemistry Chemical Physics\/} {\bf 24}
  13941--13950

\bibitem{janczura2021identifying}
Janczura J, Balcerek M, Burnecki K, Sabri A, Weiss M and Krapf D 2021 {\em New
  Journal of Physics\/} {\bf 23} 053018

\bibitem{dieball2022scattering}
Dieball C, Krapf D, Weiss M and Godec A 2022 {\em New Journal of Physics\/}
  {\bf 24} 023004

\bibitem{savin2005static}
Savin T and Doyle P~S 2005 {\em Biophysical Journal\/} {\bf 88} 623--638

\bibitem{lanoiselee2017unraveling}
Lanoisel{\'e}e Y and Grebenkov D~S 2017 {\em Physical Review E\/} {\bf 96}
  022144

\bibitem{wagner2017classification}
Wagner T, Kroll A, Haramagatti C~R, Lipinski H~G and Wiemann M 2017 {\em PLoS
  ONE\/} {\bf 12} e0170165

\bibitem{reverey2015superdiffusion}
Reverey J~F, Jeon J~H, Bao H, Leippe M, Metzler R and Selhuber-Unkel C 2015
  {\em Scientific Reports\/} {\bf 5} 1--14

\bibitem{romanczuk2012active}
Romanczuk P, B{\"a}r M, Ebeling W, Lindner B and Schimansky-Geier L 2012 {\em
  The European Physical Journal Special Topics\/} {\bf 202} 1--162

\bibitem{lemaitre2023non}
Lemaitre E, Sokolov I~M, Metzler R and Chechkin A~V 2023 {\em New Journal of
  Physics\/} {\bf 25} 013010

\end{thebibliography}
\end{document}